\documentclass[twocolumn,showpacs,preprintnumbers]{revtex4}
\usepackage{amssymb}
\usepackage{amsmath}
\usepackage{dcolumn}
\usepackage{bm}
\usepackage{graphicx}
\begin{document}

\title{DENSITY MATRIX AND FIDELITY ESTIMATION OF MULTI-PHOTON ENTANGLEMENT VIA PHASELIFT}
\author{YIPING LU}
\affiliation{School of Physics, Beijing Institute of Technology, Beijing 100081, China}
\author{QING ZHAO}
\email{qzhaoyuping@bit.edu.cn}
\affiliation{School of Physics, Beijing Institute of Technology, Beijing 100081, China}

\begin{abstract}
The experiments of multi-photon entanglements have been made by some groups, including Pan＊s group (Ref.\cite{8photons},\cite{threephoton},\cite{13}). Obviously, the increase number of the photon would cause a dramatically increase in the dimension of the measurement matrix, which result in a great consumption of time in the measurements. From a practical view, we wish to gain the most information through as little measurements as possible for the multi-photon entanglements. The low rank matrix recovery (LRMR) provides such a possibility to resolve all the issues of the measurement matrix based on less data. In this paper, we would like to verify that whether the LRMR works for six qubits and eight photons in comparison to the data given by Pan＊s group, i.e. we input a fraction of the data to calculate all of others. Through exploring their density matrix, fidelity and visibility, we find that the results remain consistent with the data provided by Pan's group, which allows us to confirm that the LRMR can simplify experimental measurements for more photons. In particular, we find that very limited data would also give excellent support to the experiment for fidelity when low rank, pure state, sparse or position information are utilized. Our analytical calculations confirm that LRMR would generalize to multi-photon state entanglement.
\end{abstract}

\maketitle

\section{Introduction}	        
\vspace*{-0.5pt}
\noindent
 With the rapid development of photonic experiments, highly entangled multi-qubit quantum state has been created successfully \cite{severalothernatures}. The polarization is employed to manufacture Greenberger-Horne-Zeilinger(GHZ) state \cite{8photons,threephoton,3photo} and the additional momentum for each photon is introduced to create effectively ten-qubit state \cite{13}. Quantum state tomography (QST) is an indispensable way to deduce unknown state from measurement of a quantum system \cite{statetomography}. Its duty is to obtain the density matrix of the prepared quantum state from a series of the positive operators valued measurements(POVM). Since the distributions of outcome frequency of any finite copies of the same measurement generally deviate from their asymptotic limits \cite{15OFULR}, we can never get the actual probabilities with any finite copies of measurements. However, the frequency converges to the true one with increasing copy number of measurements on a set of complete bases. Therefore, the precise description of realistic state can be approximately achieved by reconstructing the quantum state with limited copies \cite{statetomography}. The statistical results of measurements in the observable eigen bases specifically correspond to diagonal elements of the density matrix \cite{scirep}. Standard quantum state tomography (SQST) requires measurement on at least $d^{2}$ bases to determine all the elements of a $d \times d$ density matrix, where $d$ equals to $2^{n}$ for a n-qubit photonic system. Obviously, SQST leads to exponential increase of the number of measurements with the growth of the state space dimension, which cost considerable resources to satisfy the POVM requirement. Even if all the data from complete measurement is collected, it is still intractable to invert the frequencies to the state estimation without efficient method. Furthermore, the state description from unavoidable noisy may lead to insignificant result, such as $Tr(\rho^{2})>1$ ($\rho$ is the estimated density matrix) \cite{qubitQST}.

 Recent years, efficient quantum state tomography arouses widely interest to overcome these daunting obstacles. For matrix product state (MPS) and some other specially constructed states, partial parameters, with the number much less than $d^{2}$, are enough to embody these states\cite{1,2,4}. For general states, the mutually unbiased bases (MUB) can give a complete description with the minimum number of measurements \cite{MUB}. Its efficiency has been confirmed by the reconstruction of two entangled photons in experiment \cite{3}. Except MUB, the method of maximum entropy can also give approximate estimation of quantum state \cite{statetomographMaxEntropy}.

 In addition to these methods, a new theory, the compressed sensing (CS) \cite{cs1,cs2,5,6} is proposed by Donoho, Candes, and Tao in 2005. It can recover whole signal precisely from much less measurements, if the original signal is sparse or can be transformed into sparse one. The developed theory of CS, low-rank matrix reconstruction (LRMR) verifies the particular recovery of the original matrix \cite{MC,LRMR}. If the measurement matrix satisfies Restricted Isometry Principle (RIP) \cite{decoding,21,22}or is randomly chosen from various matrix bases, LRMR can give a good performance by solving a convex problem. This approach has been applied to deal with the density matrix for pure or near-pure states in the quantum physics \cite{DavidGross}, and further substantiated by photonic experiment\cite{2012zhongkeda}. Anyhow, no matter whether the measurement matrix satisfies the two conditions, as long as you know some characteristics of the matrix, such as rank is 1, especially each measurement matrix can be decomposed into the Kronecker product of a complex column vector and its conjugate transpose vector, then  is it still possible to recover the original matrix efficiently through the LRMR? The PhaseLift method gives the positive answer in theory when the measurement vector, which constructs the measurement matrix, is normal distribution or sphere uniform distribution \cite{phaselift1,phaselift2,phaselift3,phaselift4}. Furthermore, whether this theory can be efficiently extended to the specially constructed measurement vector based on multi-qubit entangled quantum states, or applied to the fidelity estimation and density matrix reconstruction \cite{Direct Fidelity Estimation from Few Pauli Measurements} even under lower sampling.

 In this paper, the quantum state of six-qubit Schrodinger cat (SC) is studied by applying LRMR to special case or phaselift method to process the experimental data to recover its density matrix. Specifically, the ideal SC state is pure, but a produced state in experiment would be a pure state with noise, which corresponds to a density matrix with rank 1 and added noise. Therefore, according to the experimental measurement, LRMR in the case of rank one or phaselift is exploited to calculate the density matrix from highly incomplete experimental measurement. The corresponding measurement matrices come from the matrices decomposed by entanglement witness of SC states \cite{13}. The three properties of density matrices are taken as additional constraints for the optimal method. The overall computation results are constructed density matrices, which have similar fidelities as the experimental one. They have relatively small Mean Square Error (MSE). Moreover, through reducing the number of measurement matrices, reconstruction is still performed very well, which can be revealed by the fidelity and MSE. In particular, we find 500 of measurement matrices are sufficient to recover the density matrix, which is much smaller than the number 4096 to determine all the elements of the density matrix by SQST.

\section{Experiment and method}
\subsection{Description of preparation system for six qubit SC state }

 Pan's group designed experimental setup of multi particle entanglement, and efficiently realized creation of hyper-entangled multi-qubit Schrdinger cat states\cite{8photons, threephoton, 13}. Here the experiment of six-qubit SC state will be simply introduced about the measurement matrix for phaselift. Its detailed procedure can be found in Ref.\cite{13}.

 The six-qubit SC state is the superposition of two maximal orthogonal quantum states. Its form can be expressed as
\begin{equation}
|SC\rangle=(|H\rangle^{\otimes 3}|H^{'}\rangle^{\otimes
3}+|V\rangle^{\otimes 3}|V^{'}\rangle^{\otimes 3})/\sqrt{2},
\end{equation}
where $3$ is the number of entangled photons, $H$ and $V$ denote to
horizontal and vertical polarization respectively, $H^{'}$ and
$V^{'}$ express two orthogonal spatial modes of the
photons \cite{13}. The ideal density matrix of SC state has four same numbers " 1/2 " at the four
corners while all of the other elements are zero.

The first step to prepare $|SC\rangle$ is to produce three-qubit
polarization-entangling state
$|SC\rangle^{3}_{p}=(|H\rangle^{\otimes 3}+|V\rangle^{\otimes
3})/\sqrt{2}$. The state
$(|H\rangle|H\rangle+|V\rangle|V\rangle)/\sqrt{2}$ is prepared through
spontaneous parametric down-conversion on one
path. Simultaneously, the polarization state
$(|H\rangle+|V\rangle)/\sqrt{2}$ is produced by a single photon source on the other path. Adjusting the two path meets at a point, and the delay of
photons between the two paths is eliminated through fine adjustments
to ensure the photons arriving at the polarizing beam-splitters
(PBS) simultaneously. Then, the three photons' SC state
with polarization is successfully created in the experiment.

The six-qubit SC state is produced by planting spatial modes on the
polarization-encoded three-qubit state. According to different
polarization, PBS separates the freedom of spatial modes of photons
into two orthogonal modes $H^{'}$ and $V^{'}$. State $\alpha|H\rangle|H\rangle|H\rangle|H^{'}\rangle|H^{'}\rangle|H^{'}\rangle+\beta|V\rangle|V\rangle|V\rangle|V^{'}\rangle|V^{'}\rangle|V^{'}\rangle$ is produced by employing three PBS. Finally, the completely created six-qubit SC state in experiment is obtained.

From Ref.\cite{13}, measurements are performed on a special group of bases,
so that visibility and the fidelity with the ideal SC state can be
calculated out. To be more precise, we denote $|H\rangle$=$|H'\rangle$ as logic $|0\rangle$
and $|V\rangle$=$|V'\rangle$ as $|1\rangle$. The POVM on the bases
of $|0\rangle$/$|1\rangle$, $|\pm,\theta\rangle=|0\rangle \pm e^{i\theta}|1\rangle$, are
carried out independently and simultaneously for the information of
polarization and spatial qubits. Optical interferometer concretely combines the two paths into a
non-polarizing beam splitter (NBS), and shifts the relative phase
$\theta$ by suitable delay on one of the paths. This gives the measurement on different bases $|\pm,\theta\rangle$. In addition, to avoid noise interfering one another, measuring the created state
needs Sagnac-like interferometer to overcome the instability of path
length. Lastly, state information residing on qubit of spatial mode is coherently measured by
Mach-Zehnder-type interferometer with a NBS. Accordingly, the polarization
of the state is extracted by a combination of a
quarter-wave plate, a half-wave plate, a PBS and single-photon
detectors. Therefore, the measurement of six-qubit
SC state is accomplished \cite{13}.

 The six-qubit SC state created experimentally is described by a $64\times64$ density matrix $\rho_{exp}$. Its fidelity between the ideal state and the experimental one is defined by
\begin{equation}
F_{exp}(|SC\rangle)=\langle SC|\rho_{exp}|SC\rangle=Tr(\rho_{exp}|SC\rangle\langle SC|)=1/2-\langle\textit{w}\rangle,
\label{Fexp}
\end{equation}
in which $\langle w\rangle$ is the expectation of entanglement witness
of the created SC state. We can calculate
$F_{exp}(|SC\rangle)$ by $\langle w\rangle$. Eq.(\ref{Fexp}) can be changed into the form:
\begin{equation}
\langle\textit{w}\rangle=Tr(\rho_{exp} \textit{w})=Tr(\rho_{exp}(1/2I-|SC\rangle\langle SC|))
\end{equation}
in which, $|SC\rangle\langle SC|$ is decomposed as
\begin{equation}
\begin{split}
&|SC\rangle\langle SC|\\
&=1/2[(|H\rangle\langle H|)^{\otimes 6}+(|V\rangle\langle V|)^{\otimes 6}+(1/6)\sum(-1)^{k}M_{k\pi/6}^{\otimes 6}] \label{entanglementde}
\end{split}
\end{equation}
where
$M_{k\pi/6}=cos(k\pi/6)\sigma_{x}+sin(k\pi/6)\sigma_{y}$ \cite{13,toolbox}. Prof Pan's group obtained the experimental
coincidence counts, which corresponds to all the different matrix in Eq.(\ref{entanglementde}), hence the expectations of these matrices can be calculated out. Evidently, the $F_{exp}$ is calculated as: $0.6308\pm0.0015$  \cite{13} (see appendix for details.).

Besides the fidelity estimation, usually there is no way to estimate its density matrix precisely. Since the realization of SQST will need at least $d^{2}$ various measurements for multi-quit system, while the experimental measurement is not complete. However, when the measurement data for the state is just enough to calculate the fidelity through the bridge of entanglement witness and visibility, we will show that it's also sufficiently for phaselift to determine the density matrix. Afterward, the estimation of the density matrix is described explicitly by phaselift through the partial measurements.

\subsection{Optimization of density matrix}

 Phaselift is described as following: let $\rho$ be some unknown matrix of dimension $d$ and can be decomposed into a matrix with rank $1$ , namely,
 \begin{equation}
\rho =xx^{*}, \label{rou_x}
\end{equation}
in which $x$ represents a $d \times 1$ complex vector, and $*$ denotes to conjugate transpose. Let $M_1$, $M_2$, $\cdots$, $M_i$, $\cdots$, $M_{Num}$ be a set of measurement matrices, which can also be decomposed into
  \begin{equation}
M_i=z_iz_i^{*} \label{M_v}
\end{equation}
  ; then $\rho$ can be recovered from its inner products
  \begin{equation}
   Tr(\rho M_{i})=|\langle z_i,x\rangle|^2,i=1,2,\cdots,Num?
   \end{equation}

To answer this question, rank of $\rho$, measurement matrices and properties of $\rho$, all of them need to be considered. Since the rank of the density matrix of ideal SC state is $1$, so it is suitable to relax this condition to suppose that the rank of the density matrix of state achieved in experiment is 1, which is admixed with noise. In the constraint, the rank of $(\rho)$ equals to 1, which is revealed in (\ref{rou_x}). Based on the experiment, a specific set of measurement bases is utilized, which consists of matrix $(|+,\theta\rangle\langle+,\theta|)^{\otimes 6}$, $(|+,\theta\rangle\langle+,\theta|)^{\otimes 5}(|-,\theta\rangle\langle-,\theta|)$, $\cdots$, $(|-,\theta\rangle\langle-,\theta|)^{\otimes 6}$, ($\theta=k\pi/6$, $k=0,1/4,2/4,\cdots,47/4$) plus $(|H\rangle\langle H|)^{\otimes6}$, $(|H\rangle\langle H|)^{\otimes 5}(|V\rangle\langle V|)$, $\cdots$, $(|V\rangle\langle V|)^{\otimes 6}$. Therefore, the total number of measurement matrices is $49\times 2^{6}$ = $3136$. It is noted that all of these matrices are linearly independent and can be expressed as $z_iz_i^{*}$ (See appendices for details). All $M_{i}$ are known measurement matrix acting on an unknown density matrix $\rho$ to yield a set of measurements $b_{i}$, from which $\rho$ must be recovered. In the experiment, frequencies are taken as the observed values $b_{i}$. To reconstruct density matrix, all or parts of them are selected as measurement matrices in different sampling rates. From all these selected quadratic measurements, the density matrix can be calculated. In practice, due to experimental imperfections, linear inversion may lead to the solution with no physical meaning. This issue can be circumvented by imposing the properties of the density matrix: $\rho=\rho^{H}, Tr(\rho)=1, \rho\succeq0$ to bound the
solution to the ensemble hermitian, trace 1 and positive
semidefinite in phaselift optimization \cite{2012zhongkeda}.

Considering total error of all POVM, the measurement
constraint can also be especially changed into the form of liner absolute value. Then, it
is natural to consist the problem based on Ref. \cite{cs1,DavidGross,2012zhongkeda,phaselift1,phaselift2,phaselift3,phaselift4}
 \begin{equation}
min \quad \Sigma^{Num}_{i=1}|Tr(M_{i}\rho)-b_{i}|, \quad
s.\ t. \quad \rho=\rho^{H},Tr(\rho)=1,\rho\succeq0,  \label{minimizetr2}
\end{equation}
where $M_{i}$ is defined as the measurement matrix, its number is essentially represented as sampling number $Num$, $Num/64^{2}$ is the sampling rate, $b_{i}$ denotes frequency or the measuring
value of $M_{i}$. We hope the summation of the absolute value of frequency difference
between the experiment and the construction, is the smallest, so that the reconstructed density matrix could be the best one fitting the experiment. Eq.(\ref{minimizetr2}) is sufficient to yield the density matrix $\rho$, the
optimal approximation to the state $\rho_{exp}$.

\section{Results and discussions}
\subsection{Calculations of the density matrix for six qubit SC state}

 The experimental data admixed with noise, i.e, $b_{i}$, ($i=1,\cdots,Num$), is supplied by Pan's group \cite{13}. Based on the measurement and method illustrated in the previous section, Eq.(\ref{minimizetr2}) is employed to calculate the density matrix of six-qubit SC state. In this section, we focus on the results of concrete construction. Firstly, numerical simulation is conducted, and the ideal constructed density matrix is obtained. Subsequently, the calculation of the density matrix from experimental data is carried out through phaselift. Lastly, recovery of the density matrix from measurement data with the minimum number (around 500) is performed, which sufficiently confirm high-quality estimation of fidelity and quantitative MSE. Besides, the problem in different case are all calculated by cvx \cite{cvx}.

Computer simulation with the ideal SC system is carried out to verify the reliability of Eq.(\ref{minimizetr2}). The density matrix of ideal SC is assumed as the matrix being observed and the measurement matrices "$M_{i}$" is obtained from an entanglement witness of
SC (GHZ) state described before \cite{toolbox}, and the simulated data is obtained by $Tr(M_{i}|SC\rangle\langle SC|)$. The noise is free under this ideal situation. Through executing (\ref{minimizetr2}), density matrix $\rho_{simu}$ is constructed;
the corresponding fidelity is 1, and its error is equal to zero: $||\rho_{simu}-|SC\rangle\langle
SC|||_{F}=0$.

\begin{figure}[tbp]
\centering
\includegraphics[width=0.67\columnwidth]{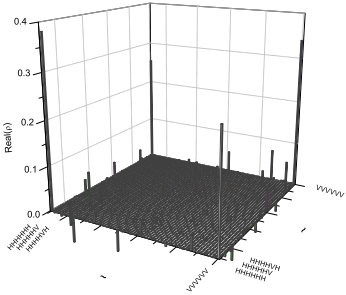}~~~%
\caption{(Color online). Real parts of $\rho_{3136}$. The result is obtained by constructing method (\ref{minimizetr2}), with sampling rate $(49\times2^6)/(2^{6})^{2}$$\approx$$0.7656$.
The real parts of two large elements on the diagonal of the density matrix
equal to 0.3816 on $|HHHHHH\rangle\langle HHHHHH|$ and 0.3402 on
$|VVVVVV\rangle\langle VVVVVV|$ respectively. And the two main
elements on the anti-diagonal are both 0.2544 on $|HHHHHH\rangle\langle
VVVVVV|$ and $|VVVVVV\rangle\langle HHHHHH|$. \label{Fig49_64}}
\end{figure}
\begin{figure}[tbp]
\centering
\includegraphics[width=0.67\columnwidth]{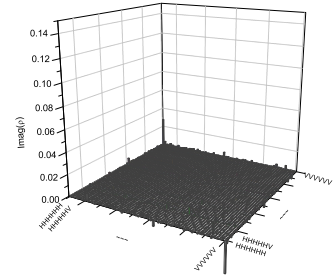}
\caption{(Color online). Imaginary parts of $\rho_{3136}$. The height of the pillar
represents the numerical value of the elements of $\rho_{3136}$ obtained by
solving (\ref{minimizetr2}), with sampling rate
$(49\times2^6)/(2^{6})^{2}$$\approx$$0.7656$. \label{Fig49_64_IMAG}}
\end{figure}

\begin{figure}[tbp]
\centering
\includegraphics[width=0.87\columnwidth]{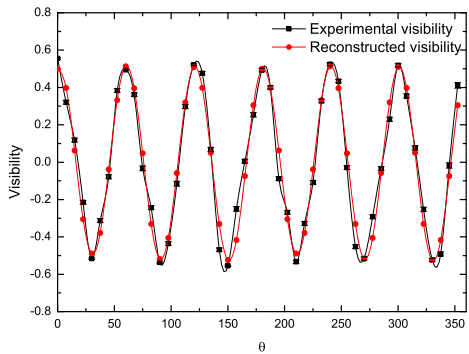}~~~%
\caption{(Color online). Constructed visibility with experiment and
reconstruction. Namely, the value $\langle M_{\theta}^{\otimes 6}\rangle$ in different cases. Black
points are the experimental visibility obtained from Ref.\cite{13}. Red points are the visibility i.e, $\langle M_{\theta}^{\otimes 6}\rangle=Tr(M_{\theta}^{\otimes
6} \rho_{3136})$ for
$\theta={k\pi/6}, k=0,1/4,2/4,\cdots,11$. The
lines are draw based on the data points. When $\theta$ are $0$, $\pi/6$, $\pi/3$, $\cdots$,
$\langle M_{\theta}^{\otimes 6}\rangle$ correspond to the summit or
trough, which are in good agreement with the experiment. When
$\theta$ equals to other values, $\langle M_{\theta}^{\otimes
6}\rangle$ deviate the experimental ones a little bit, which are
caused by the inexact construction of density matrix. \label{visibility4964} }
\end{figure}

In Pan's experiment, all the statistical results of coincidence counts are fully
collected on different bases \cite{13}. To obtain the precise density matrix
agreeable with them, Eq.(\ref{minimizetr2}) is solved  by employing all the data
measured in the experiment, i.e $3136$ in all. The
result $\rho_{3136}$ is shown in Fig.\ref{Fig49_64} and Fig.\ref{Fig49_64_IMAG}. To identify how close the
constructed state is to the aimed one $|SC\rangle=1/\sqrt{2}[|1\rangle^{\otimes6}+|0\rangle^{\otimes6}]$, the extent of
overlap between them is given by $F_{3136}=\langle SC|\rho_{3136}|SC\rangle$, which is $0.6154$. Moreover, the corresponding visibility is also shown in Fig.\ref{visibility4964}, which nearly fits the results given in experiment.

\begin{figure}[tbp]
\centering
\includegraphics[width=0.67\columnwidth]{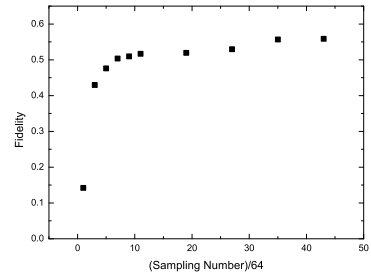}~~~%
\caption{(Color online). The picture of fidelity under different samplings. Each point is gain by averaging the 12 values gain under the same sampling. \label{fidelity}}
\end{figure}

 \begin{figure}[tbp]
\centering
\includegraphics[width=0.67\columnwidth]{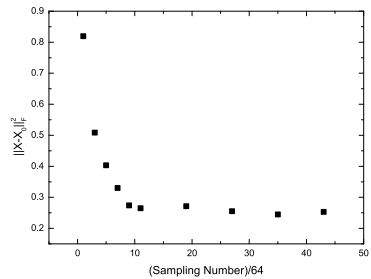}~~~%
\caption{(Color online). The picture of error under different samplings. When the sampling number is larger than $64\times8$ (around 500), the error tends to be a constant.  \label{error}}
\end{figure}

The similar results can also be derived by less data. Parts of the measurement matrixes are randomly selected under different samplings without replacement. By solving Eq.(\ref{minimizetr2}) repeatedly in various of sampling, the density matrices are reconstructed. The fidelities with ideal state are shown in Fig.\ref{fidelity}, and MSE is presented in Fig.\ref{error}.

The above remarkable results can also be derived by choosing even less data. According to the definition
$F=Tr(\rho|SC\rangle\langle SC|)$, only
$|0\rangle\langle0|^{\otimes6}$, $|1\rangle\langle1|^{\otimes6}$,
$M_{\theta}^{\otimes6}=(cos\theta \sigma_{x}+sin\theta
\sigma_{y})^{\otimes6},\theta=0,1/12\cdot2\pi,2/12\cdot2\pi,\cdots,11/12\cdot2\pi$ are specifically chosen, which ensures that measurement matrices have nonzero elements in the four corners of experimental density matrix (positions (1,1),(1,64),(64,1) and (64,64)). By solving Eq.(\ref{minimizetr2}), similar fidelity is obtained for this situation ($0.625$). This result reveals that with severe
undersampling, the constructed density matrix still has similar fidelity with
the established one under high samplings when the priori knowledge of the nonzero positions in the density matrix of SC state is applied.

\begin{figure}[tbp]
\centering
\includegraphics[width=0.67\columnwidth]{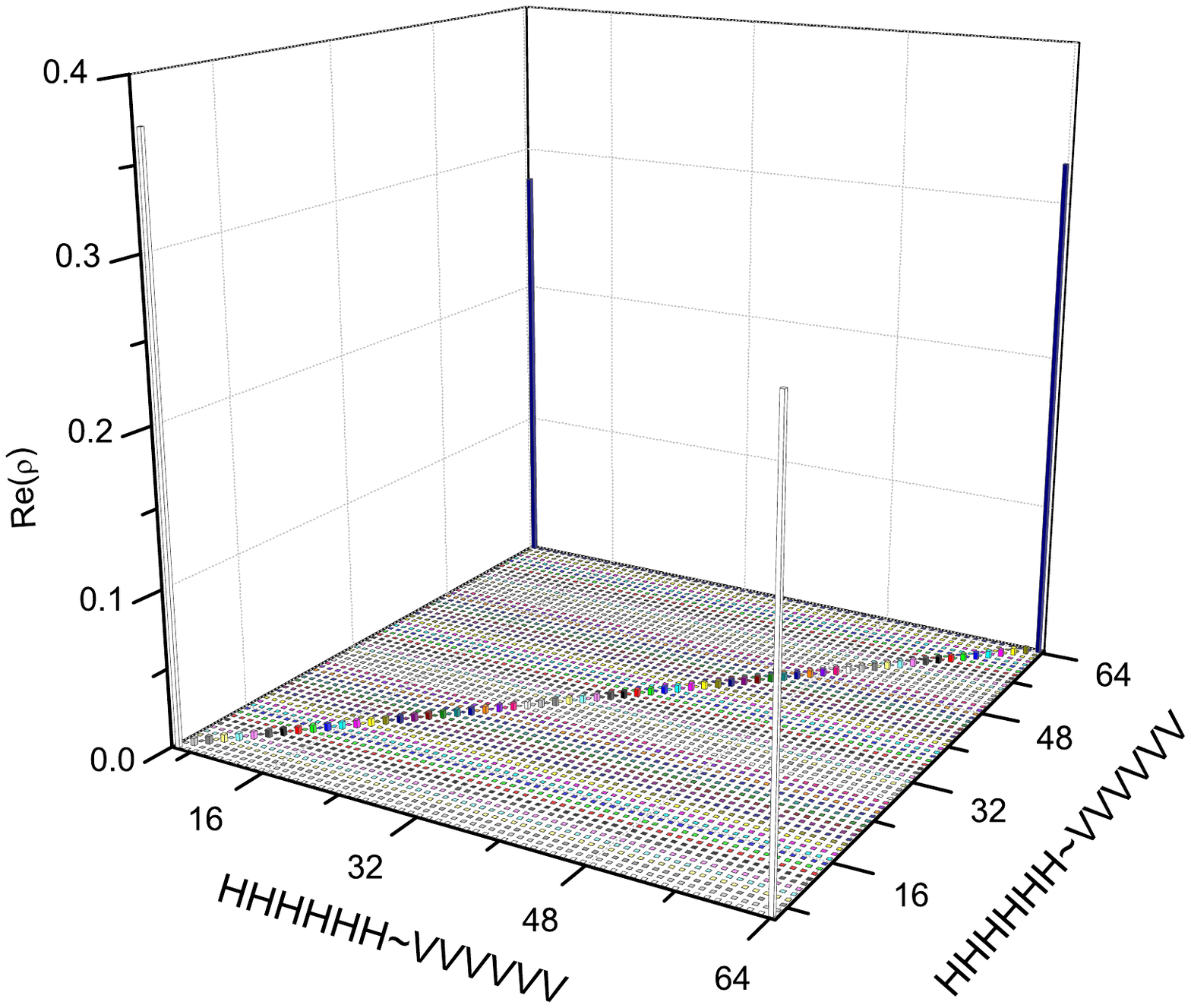}~~~%
\caption{(Color online). Real parts of constructed density matrix $\rho_{3}$ for
six-qubit SC state. In this case, sampling number is three (sampling rate = $3/(2^6)^{2}$) and these 3 measurement matrices are specially chosen, i.e.
$|0\rangle^{\otimes6}\langle0|^{\otimes6}$, $|1\rangle^{\otimes6}\langle1|^{\otimes6}$ and
$M_{\theta}^{\otimes6} (\theta=0)$. When (\ref{minimizetr2}) is implemented, $\rho_{3}$ is drawn, in which the two
large elements are on (1,1), (64,64), corresponding to diagonal
elements of $|HHHHHH\rangle$ and $|VVVVVV\rangle$, along with large
positive values on (1,64), (64,1) indicate that this constructed
state has the qualities of the desired SC state. The imaginary parts of all elements of
constructed density matrix are near zero (average absolute values $<$ 0.001), so are not shown. \label{routr3}}
\end{figure}

\begin{figure}[tbp]
\centering
\includegraphics[width=0.87\columnwidth]{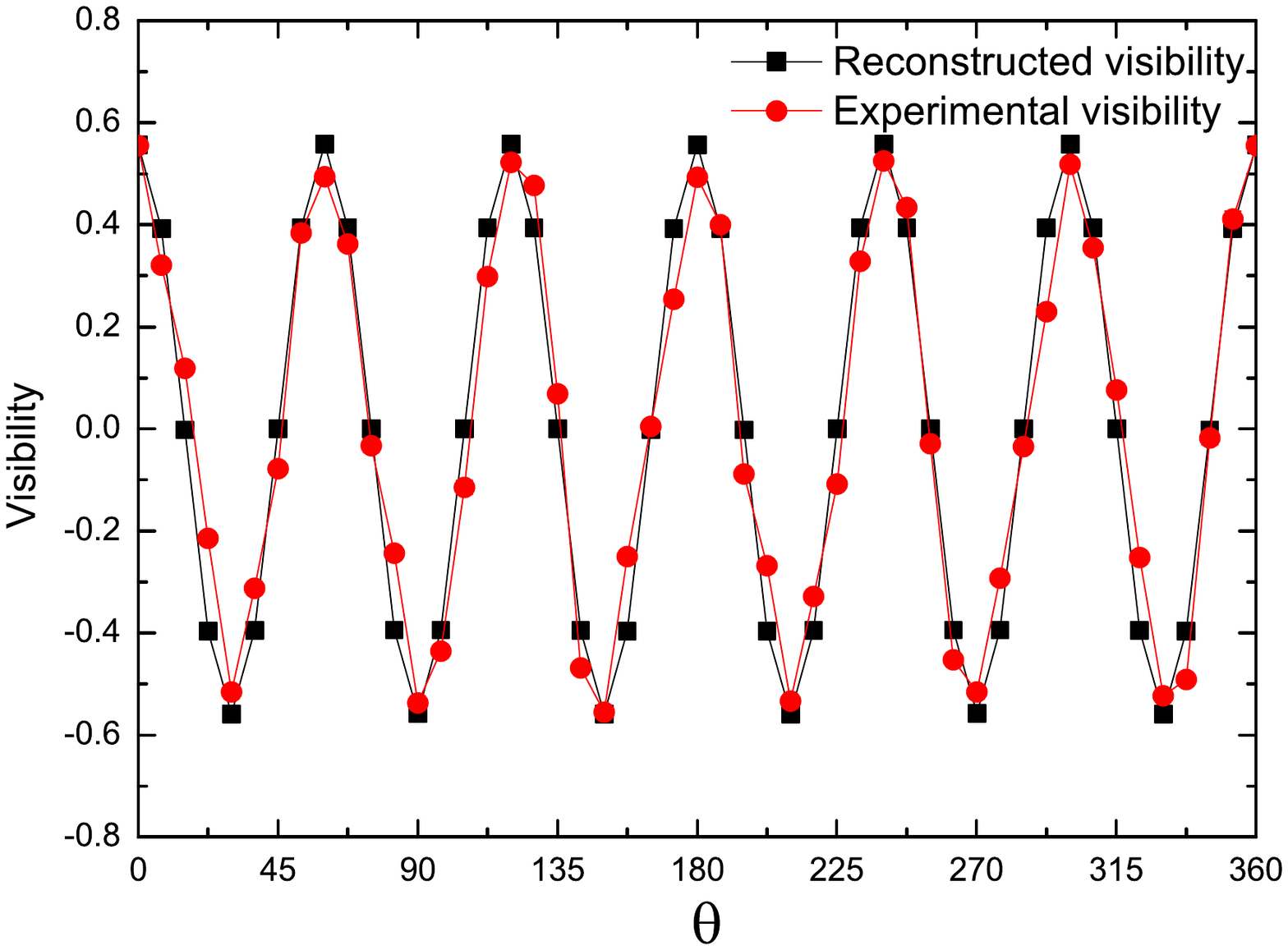}~~~%
\caption{(Color online). Sketch of constructed visibility or the match between experiment and
reconstruction for $\langle M_{\theta}^{\otimes 6}\rangle$. The lines are draw based on the data points, where the red ones are obtained from the experiment \cite{13} and the black ones are the expectations of
different $M_{\theta}^{\otimes 6}$ obtained by solving (\ref{minimizetr2}) when sampling number is three. The values, $\langle M_{\theta}^{\otimes 6}\rangle=Tr(M_{\theta}^{\otimes
6}\rho_{3})$, when $\theta$ are $0$, $\pi/6$, $\pi/3$, $\cdots$,
$\langle M_{\theta}^{\otimes 6}\rangle$ correspond to the summit or
trough, which are in good agreement with the experiment. When
$\theta$ equals to other values, $\langle M_{\theta}^{\otimes
6}\rangle$ deviate the experimental ones a little bit, which are
caused by the low sampling rate and the inexact construction of density matrix. \label{motion2} }
\end{figure}

Subsequently, further reducing the sampling number to three is proceeded similarly, which is sufficient to estimate fidelity too. Intuitively, only one off-diagonal measurement element $M_{\theta}^{\otimes6}$ ($\theta$=$0$) and two largest elements
corresponding to $|HHHHHH\rangle$ and $|VVVVVV\rangle$ or the first and last elements on the
diagonal of the density matrix $|0\rangle\langle0|^{\otimes6}$,
$|1\rangle\langle1|^{\otimes6}$ are selected. Then through solving Eq.(\ref{minimizetr2}), $\rho_{3}$ is gained, and its result is shown in Fig.\ref{routr3}. Its fidelity with the ideal SC state is $F_{3}=\langle
SC|\rho_{3}|SC\rangle=0.6274$, which is almost same as the one of Pan's group
$(0.6308)$. The corresponding visibility is also shown in Fig.\ref{motion2}. However, by comparing the results in Fig.\ref{routr3}, Fig.\ref{Fig49_64} and Fig.\ref{Fig49_64_IMAG}, it is not necessary to rely on the fidelity to certify that the state reconstruction is well performed by Phaselift.

\subsection{Discussions}

  Obviously, the prior knowledge of rank 1, is required for constructing the density matrix for six-qubit SC state, and the information of positions for the four largest modulus elements of the density matrix is also needed according to the formula of fidelity estimation when choosing very limited measurement matrices to estimate fidelity. For this reason, the fidelity estimations of the density matrix by PhaseLift perform very well from experimental data. More generally, the method may apply to other entangled states too, such as W state, C state \cite{othernature}. Overall, the fidelity estimation is quite good, while it is still required to study in the future that if the measurement matrices satisfy the RIP, and the error bound also needs to be estimated.

  Up to now we have focused on using PhaseLift to solve the state estimation. Most elements in Fig.\ref{routr3} approach to zero, which indicates that sparse characteristic might be used to estimate fidelity too. It is interesting to find that an alternate scenario is $L_{1}$ optimization \cite{L0L1}. Especially since there are many zero elements in the density matrices of ideal GHZ (SC) state, W state and C state \cite{othernature}, thus the one achieved in experiment might be seen as a sparse matrix too. If its density matrix is taken as a sparse vector, constructing the density matrix
means to recover the sparse vector, which is a $L_{0}$ optimization
problem. However, it is N-P hard \cite{NPhard} problem, and can be validly replaced by
$L_{1}$ optimization \cite{L0L1}. Particularly, for an ideal SC state, its density matrix is
a sparse matrix with only four non-zero elements; Therefore, $L_{1}$
optimization could reproduce the experimental results. Under this
ansatz, in the case of the three constraints of $\rho$, we also
apply this method to construct experimental six
qubit SC state. Because the object of $L_{1}$ optimization is a
vector, matrix $\rho$ is transformed into vector
$\overrightarrow{\rho}$, a column vector connecting all
the columns of matrix $\rho$ from left to right one by one. From the previous analysis and all constraints of $\rho$, the optimization problem
 \begin{equation}
  \begin{split}
&min||\overrightarrow{\rho}||_{1} \quad\\
&s.\ t. Tr(M_{i}\rho)=b_{i}, i=1,\cdots Num.\  Tr(\rho)=1,\ \rho^{H}=\rho,\ \rho\succeq0,  \label{L1con}
 \end{split}
 \end{equation}
is constructed, and its solution is defined as $\rho_{L1}$.
Interestingly, when three same measurement operators
$|0\rangle\langle0|^{\otimes6}$, $|1\rangle\langle1|^{\otimes6}$,
$M_{\theta}^{\otimes6}$ ($\theta=0$) are chosen for
$M_{i}$($i=1,2,3.$) in Eq.(\ref{L1con}), the density matrix of
six-qubit experimental SC state $\rho_{L1}$ can be obtained evidently, which has similar fidelity with the real one, but different elements' distribution in the density matrix. The fidelity between
the reconstructed density matrix $\rho_{L1}$ and the ideal density
matrix $|SC\rangle\langle SC|$ is 0.6411, similar to the
fidelity 0.6308 obtained from experiment data directly, which gives some insight into the optimal measurement design of experiment for fidelity estimation.

  Therefore, fidelity of the state with sparse characteristic might be estimated by $L_{1}$ optimization. Pure or pure states mixed with noise can be reconstructed by PhaseLift. Since experimental SC states fulfill both conditions, hence, $L_{1}$ optimization and PhaseLift perform excellently for fidelity estimation even with extremely limited measurements.

  Further more, we also calculated the entropy for the density matrix we get, which is $0.7016$ according to the definition: $-Tr(\rho_{3136} log_e \rho_{3136})$ \cite{entropy1}.

\section{Eight photon density matrix and fidelity estimation}
\begin{figure}
\centering
\includegraphics[width=1.2\columnwidth]{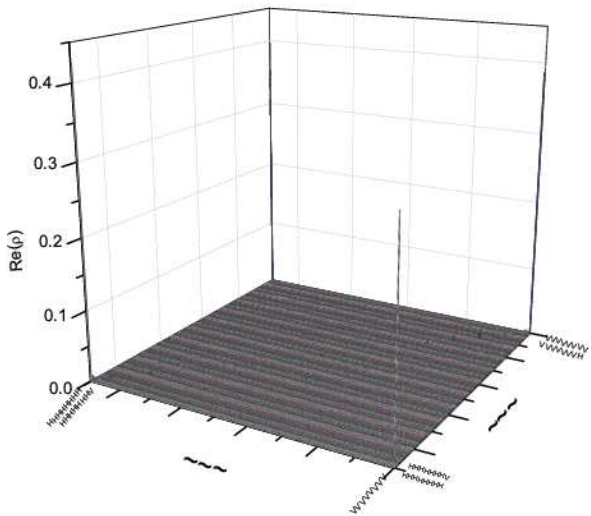}~~~%
\caption{(Color online). Constructed experimental density matrix of eight photon SC state. The value of the elements on $(1,1)$ or $|H\rangle^{\otimes 8} \langle H|^{\otimes 8}$ is $0.4206$, and $0.3864$ for $|V\rangle^{\otimes 8} \langle V|^{\otimes 8}$, $0.3043$ for $|H\rangle^{\otimes 8} \langle V|^{\otimes 8}$ and $|V\rangle^{\otimes 8} \langle H|^{\otimes 8}$. All the absolute value of image part of the elements are less than $10^{-7}$, so are not drawn. So the fidelity is $0.7078$, which is consistent with the $0.708\pm0.016$ directly given by the eight-photon experiment.\label{8photonDM}}
\end{figure}

Additionally, based on the Ref.\cite{8photons} and the corresponding data given by Pan's group, the eight photon density matrix can also be calculated by solving Eq.(\ref{minimizetr2}). However, $M_i$ can not be decomposed as $z_iz_i^*$ for the given data , and its result is presented in Fig.\ref{8photonDM}.

For fidelity estimation, since the desired density matrix of SC state has only four elements at the four corners belong to nonzero; we take the position information as the priori knowledge to choose three observation matrix, which could get the information of these 4 points, and then estimate the experimental fidelity. Based on the values at four corners of reconstructed matrix are $0.4205$, $0.3864$, $0.2999$ and $0.2999$. The fidelity is calculated to be $0.7034$, which is also similar with the $0.708\pm0.016$ given by the eight-photon experiment directly.

From Fig.\ref{8photonDM}, it can be seen that the reconstruction of eight-photo experimental data is better than the recovery of six-qubits. There are three main reasons. Firstly, data has higher precision in eight-photon experiment. Because there are only nine eight-photon counts per hour, it has small probability to cause wrong counts for the detector. And the experiments last longer, then the frequency obtained can be more accurate. Therefore, the noise control is improved much better, and the ratio of corresponding signal to noise reaches $530:1$ \cite{8photons}. Secondly, measurement values are acquired from interim results in eight photo case, which can avoid some errors. Precisely, observation matrixes are the matrixes corresponding to all the diagonal elements in the density matrix and $M_{k\pi/8}^{\otimes 8}$, $k=0,1,\cdots,7$, which is obtained similarly with the $M_{k\pi/6}^{\otimes 6}$ in six-qubits case. Thus, the error is lower. Finally, the fidelity is only one way to judge the effect of reconstruction, which cannot fully illustrate the effect of reconstruction. Besides, the uniqueness of the solution of Eq.(\ref{minimizetr2}) needs to be proofed under several conditions, such as specially chosen sampling matrices, sampling rate $(256+8)/(256^2)$, and priori knowledge (rank 1 and sparse). The rank of the ideal SC state density matrix of eight photons is 1, and the matrix has only four nonzero elements among $256^2$ elements.

The processing results of eight photon entanglement allow us further to see the CS' potential power to handle a large workload of calculations and measurements of more photon entanglement. Its more detailed discussion, and theoretical analysis will be appeared elsewhere.

\section{Conclusions}
\noindent

Overall, with the promise of CS \cite{cs1}, LRMR \cite{LRMR} and PhaseLift \cite{phaselift2}, the calculated density matrix is realistic in the scheme with approximately rank 1. Besides, since four elements with the largest moduli in the density matrix of SC state are effectively sampled by three or more specially chosen operators, our result can assist experiment to use part of POVM to give almost same estimation on the fidelity of created SC state. These results can be generalized beyond the class of SC state. Essentially, one can replace SC state with any rank 1 or sparse states. Briefly PhaseLift plays pivotal roles to provide efficient evidence for the application in
large number of entangling photons. It reveals its potential power in
saving considerable resources in the experiment of hyper-entangled
multi-qubit state. However, it is just a preliminary exploration of our
work in this field, more detailed study and theory will be completed
in the near future.

\section{Acknolodgement}


\noindent
The authors would like to greatly thank Prof. Jian-wei Pan, Prof. Chaoyang Lu et.al for providing all their experimental data to us and Yulong Liu for helpful discussion. This work is in part supported by NSF of China with the Grant No. 11275024. Additional support was provided by the Ministry of Science and Technology of China (2013YQ030595-3, and 2011AA120101).

\appendix

\noindent

\noindent For six-qubit SC state, $n=6$, the expectation of the
third term on the right hand of the Eq.(\ref{entanglementde}) is
\begin{equation}
\begin{split}
&1/2[1/6\Sigma_{k=1}^{n}Tr(\rho_{exp}(-1)^{k}M_{k\pi/6}^{\otimes 6})]\\
&=1/12[-Tr(\rho_{exp}M_{\pi/6}^{\otimes 6})+Tr(\rho_{exp}M_{2\pi/6}^{\otimes 6})-Tr(\rho_{exp}M_{3\pi/6}^{\otimes 6})\\
&+Tr(\rho_{exp}M_{4\pi/6}^{\otimes 6})-Tr(\rho_{exp}M_{5\pi/6}^{\otimes 6})+Tr(\rho_{exp}M_{\pi}^{\otimes 6})]\\
&=1/12\sum_{k=1}^{6}(-1)^{k}\langle M_{k\pi/6}^{\otimes 6}\rangle,
\end{split}
\end{equation}
 in which $\langle M_{k\pi/6}^{\otimes 6}\rangle$ represents the expectation of operator $M_{k\pi/6}^{\otimes 6}$.

  It is necessary to count $\langle\textit{w}\rangle$ to calculate $F_{exp}$, which means each term decomposed by $|SC\rangle\langle SC|$ has to be measured for their expectations. More precisely, measurements on different combination of $|H\rangle$, $|V\rangle$ bases are to verify most coincidence events belong to $|HHHHHH\rangle$ and $|VVVVVV\rangle$, and measurements on $|+,\theta\rangle=|0\rangle+e^{i\theta}|1\rangle$ and $|-,\theta\rangle=|0\rangle-e^{i\theta}|1\rangle$ are to reveal photon's coherence properties for calculating $M_{\theta}^{\otimes 6}$, $\theta=k\pi/6$. Since the estimation of the expectation value of operator $M_{k\pi/n}^{\otimes6}=(|+,\theta\rangle\langle+,\theta|-|-,\theta\rangle\langle-,\theta|)^{\otimes6}$ is equivalent to the measurement on local bases ${|+,\theta\rangle,|-,\theta\rangle}$. Owing to
 \begin{equation}
 \begin{split}
&M_{k\pi/n}^{\otimes 6}\\
&=(|+,\theta\rangle\langle+,\theta|-|-,\theta\rangle\langle-,\theta|)^{\otimes 6}\\
&=(|+,\theta\rangle\langle+,\theta|)^{\otimes 6}-(|+,\theta\rangle\langle+,\theta|)^{\otimes 5}(|-,\theta\rangle\langle-,\theta|)\\
&+\cdots+(|-,\theta\rangle\langle-,\theta|)^{\otimes 6},
\label{decompose}
\end{split}
\end{equation}
  there are 64 terms in all when $\theta$ is fixed in Eq.(\ref{decompose}). When the coincidence number of photons measured on the different combination of bases $|+,\theta\rangle$, $|-,\theta\rangle$ are collected,  the number of photons corresponding to $M_{k\pi/n}^{\otimes 6}$ can be calculated from Eq.(\ref{decompose}). Thus, $\langle M_{k\pi/n}^{\otimes 6}\rangle$ can be obtained \cite{ExpWitness,13}. From these measurements, the expectations of different terms decomposed by SC state entanglement witness can be acquired.

\end{document}